\documentclass[%
 reprint,
 amsmath,amssymb,
 aps,
 prx,
superscriptaddress,
]{revtex4-2}

\usepackage{xcolor}

\usepackage{enumitem}

\usepackage{graphicx}
\usepackage{dcolumn}
\usepackage{bm}

\begin{document}

\title{First demonstration of 6\,dB quantum noise reduction in a kilometer scale gravitational wave observatory}

\author{James Lough}
\affiliation{Institut f{\"u}r Gravitationsphysik, Leibniz Universit{\"a}t Hannover and Max-Planck-Institut f{\"u}r Gravitationsphysik (Albert-Einstein-Institut), Callinstra{\ss}e 38, 30167 Hannover, Germany}
 \email{james.lough@aei.mpg.de}

\author{Emil Schreiber}
\affiliation{Institut f{\"u}r Gravitationsphysik, Leibniz Universit{\"a}t Hannover and Max-Planck-Institut f{\"u}r Gravitationsphysik (Albert-Einstein-Institut), Callinstra{\ss}e 38, 30167 Hannover, Germany}

\author{Fabio Bergamin}
\affiliation{Institut f{\"u}r Gravitationsphysik, Leibniz Universit{\"a}t Hannover and Max-Planck-Institut f{\"u}r Gravitationsphysik (Albert-Einstein-Institut), Callinstra{\ss}e 38, 30167 Hannover, Germany}

\author{Hartmut Grote}
\affiliation{School of Physics and Astronomy, Cardiff University, The Parade, CF24 3AA, United Kingdom}

\author{Moritz Mehmet}
\affiliation{Institut f{\"u}r Gravitationsphysik, Leibniz Universit{\"a}t Hannover and Max-Planck-Institut f{\"u}r Gravitationsphysik (Albert-Einstein-Institut), Callinstra{\ss}e 38, 30167 Hannover, Germany}

\author{Henning Vahlbruch}
\affiliation{Institut f{\"u}r Gravitationsphysik, Leibniz Universit{\"a}t Hannover and Max-Planck-Institut f{\"u}r Gravitationsphysik (Albert-Einstein-Institut), Callinstra{\ss}e 38, 30167 Hannover, Germany}

\author{Christoph Affeldt}
\affiliation{Institut f{\"u}r Gravitationsphysik, Leibniz Universit{\"a}t Hannover and Max-Planck-Institut f{\"u}r Gravitationsphysik (Albert-Einstein-Institut), Callinstra{\ss}e 38, 30167 Hannover, Germany}

\author{Marc Brinkmann}

\author{Aparna Bisht}
\affiliation{Institut f{\"u}r Gravitationsphysik, Leibniz Universit{\"a}t Hannover and Max-Planck-Institut f{\"u}r Gravitationsphysik (Albert-Einstein-Institut), Callinstra{\ss}e 38, 30167 Hannover, Germany}

\author{Volker Kringel}

\author{Harald L{\"u}ck}
\affiliation{Institut f{\"u}r Gravitationsphysik, Leibniz Universit{\"a}t Hannover and Max-Planck-Institut f{\"u}r Gravitationsphysik (Albert-Einstein-Institut), Callinstra{\ss}e 38, 30167 Hannover, Germany}

\author{Nikhil Mukund}
\affiliation{Institut f{\"u}r Gravitationsphysik, Leibniz Universit{\"a}t Hannover and Max-Planck-Institut f{\"u}r Gravitationsphysik (Albert-Einstein-Institut), Callinstra{\ss}e 38, 30167 Hannover, Germany}

\author{Severin Nadji}
\affiliation{Institut f{\"u}r Gravitationsphysik, Leibniz Universit{\"a}t Hannover and Max-Planck-Institut f{\"u}r Gravitationsphysik (Albert-Einstein-Institut), Callinstra{\ss}e 38, 30167 Hannover, Germany}

\author{Borja Sorazu}
\affiliation{SUPA, University of Glasgow, Glasgow G12 8QQ, United Kingdom}

\author{Kenneth Strain}
\affiliation{SUPA, University of Glasgow, Glasgow G12 8QQ, United Kingdom}

\author{Michael Weinert}

\author{Karsten Danzmann}
\affiliation{Institut f{\"u}r Gravitationsphysik, Leibniz Universit{\"a}t Hannover and Max-Planck-Institut f{\"u}r Gravitationsphysik (Albert-Einstein-Institut), Callinstra{\ss}e 38, 30167 Hannover, Germany}

\date{\today}

\begin{abstract}

Photon shot noise, arising from the quantum-mechanical nature of the light, currently limits the sensitivity of all the gravitational wave observatories at frequencies above one kilohertz.
We report a successful application of squeezed vacuum states of light at the GEO\,600 observatory and demonstrate for the first time a reduction of quantum noise up to $6.03 \pm 0.02$ dB in a kilometer-scale interferometer.
This is equivalent at high frequencies to increasing the laser power circulating in the interferometer by a factor of four.
Achieving this milestone, a key goal for the upgrades of the advanced detectors, required a better understanding of the noise sources and losses, and implementation of robust control schemes to mitigate their contributions.
In particular, we address the optical losses from beam propagation, phase noise from the squeezing ellipse, and backscattered light from the squeezed light source.
The expertise gained from this work carried out at GEO 600 provides insight towards the implementation of 10 dB of squeezing envisioned for third-generation gravitational wave detectors.

\end{abstract}

\maketitle

\section{Introduction}

  Gravitational waves, ripples in the fabric of spacetime, were long theorized, and it took
  100 years to measure them directly \cite{PhysRevLett.116.061102}.
  The currently favored devices for measuring the spacetime metric perturbations are km scale variants of the Michelson interferometer.
  These interferometers sense the differential arm length modulation due to audioband gravitational waves.

The main source of noise in all interferometric gravitational wave detectors at high frequency is the so-called shot noise, 
which can be interpreted as random detection of photons.
The quantum nature of coherent light results in uncertainty in amplitude and phase which are canonically conjugate observables. These can be transformed into the sine and cosine quadrature amplitude basis. From normalized quadrature operators, $X_1$ and $X_2$ we have the uncertainty relation,
\begin{equation}
    \label{eq:heisenberg}
    \Delta X_1 \Delta X_2 \geq 1.
\end{equation}
The coherent state is a state with minimum uncertainty equal in the two quadratures.
Reducing the uncertainty in one quadrature below this state of minimum uncertainty is known as squeezing.
Due to Heisenberg's uncertainty principle (eq.~\ref{eq:heisenberg}) this implies that the uncertainty in the orthogonal quadrature must be increased, an effect known as antisqueezing.

Quantum noise in a gravitational wave detector can be thought of as arising from vacuum fluctuations of light incident on the so-called dark port of the interferometer.
This is the output port where traditionally no light is injected and thus it is only this vacuum field that enters the output of the detector (see figure \ref{fig:aufbau}).
With this picture, the application of a squeezed vacuum field was first proposed in 1981 \cite{Caves1981Quantum-mechanicalInterferometer}.
Since then, GEO\,600 was the first km-scale detector to apply squeezing in 2010~\cite{Collaboration2011ALimit} followed by a proof of principle at the initial LIGO detector in Hanford, WA.
Long term stable application of squeezing was pioneered at GEO\,600 
\cite{Grote2013}
and continuously improved as reported here.
Recently the LIGO and Virgo detectors have reached 3\,dB
noise reduction from the application of squeezing~\cite{Tse2019,Acernese2019}.
The application of squeezed light is a key technology in the design of all third generation detectors demanding a quantum shot noise reduction of 10\,dB.

Work at the GEO\,600 gravitational wave detector focuses on instrument science research and development of technologies applied to a fully functioning gravitational wave detector.
In particular we are interested in the research and development of the application of squeezing as a technology for improving sensitivity at high frequency.
Important sources of gravitational waves at high frequency include most notably the post-merger signal around a few kHz from binary neutron star inspirals which will be signatures rich in information about the structure of neutron stars \cite{Clark2016,Abbott2019}. Additionally, at lower frequencies around 1\,kHz, tidal deformation effects at the later part of the inspiral phase that also depend on the neutron star equation of state will be visible \cite{Read2009,Chatziioannou2018}.

Additional signals of interest relevant for GEO could come from low mass dark matter fields.
It has been proposed that large interferometers could be sensitive to time varying signatures of such fields \cite{PhysRevLett.114.161301}.
The coupling comes through influence on physical constants such as the fine structure constant, which affect properties of the beamsplitter bulk material.
Due to the the topology of GEO\,600 (see figure~\ref{fig:aufbau}), the detector is highly sensitive to the beamsplitter as it is located in the most sensitive part of the interferometer
\cite{PhysRevResearch.1.033187}.

For the application of squeezed light, our goal is to reduce the uncertainty of the amplitude of light at the main read-out photodetector. 
Since this amplitude corresponds to the differential phase of light in the arms of the gravitational wave detector,
we call this phase squeezing.
At low frequencies inside the interferometer, due to the finite mass of our suspended mirrors
there is a coupling between the amplitude and phase quadratures.
However, at GEO\,600 the noise from this effect is masked by technical noise, therefore we can apply frequency independent squeezing,
phase squeezing at all frequencies, without impacting the sensitivity at low frequencies.

In this paper we present the observation of up to $6.03\,\pm\,0.02$\,dB of noise reduction at 6\,kHz due to the application of squeezing,
demonstrated for the first time on a full scale gravitational wave detector.
We establish the application of squeezing as a method for significantly reducing
the quantum noise in gravitational wave detectors equivalent to a factor of four increase
in laser power, marking a major milestone in the advancement of squeezed light application.

\section{Experimental setup}

\begin{figure}
    \includegraphics[width=\linewidth]{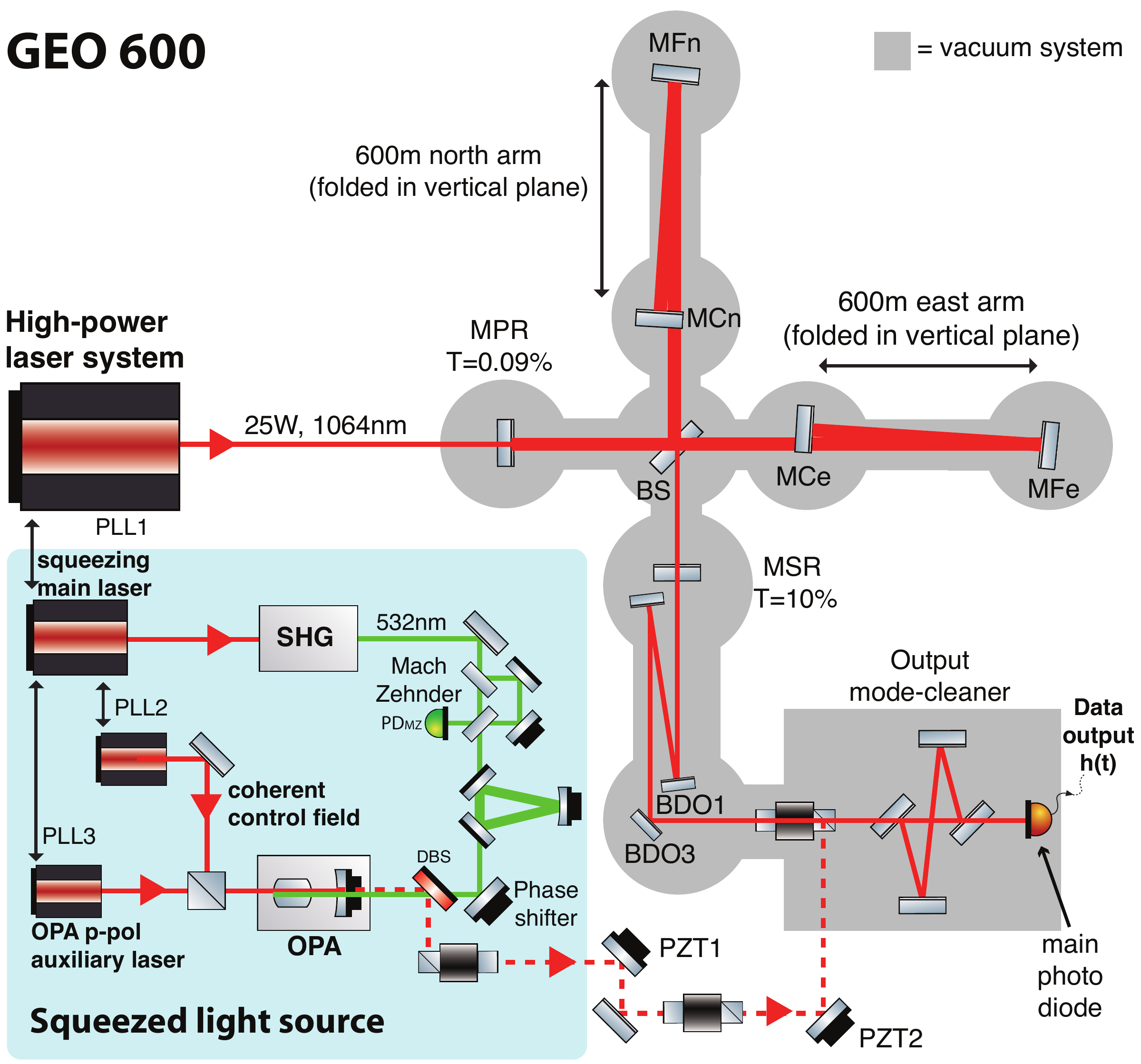}
    \caption{\label{fig:aufbau}The layout of the squeezing application to the GEO\,600 detector.
    The three Faraday isolators used for squeezed light injection are depicted.
    The squeezed light is injected such that it overlaps the main beam with the correct polarization at the MSR.
    }
\end{figure}

The GEO\,600 detector is a Michelson interferometer with power recycling and signal recycling
\cite{PhysRevD.38.2317}.
A simplified optical layout of the detector is provided in figure~\ref{fig:aufbau}.
The two arms have a folding mirror, resulting in a total length of 1200\,m per arm.
These, along with the beamsplitter mirror (BS), comprise the Michelson portion of the interferometer.
The power recycling cavity (PRC) and signal recycling cavity (SRC) are formed by the power recycling mirror (MPR) and signal recycling mirror (MSR) respectively with the Michelson.
The PRC increases the circulating power in the interferometer by a factor of about 1000 which leads to increased sensitivity.
The application of signal recycling resonantly enhances the signal.
This resonant enhancement also limits the bandwidth of the detector, acting as a low pass filter, which causes the white shot noise to increase towards higher frequency in the calibrated strain data (see figure~\ref{fig:sqzref}).
The signal due to differential arm length changes is imprinted onto the carrier light transmitted by the signal-recycling mirror.
This field is filtered with an output mode cleaner cavity (OMC), to suppress transverse and longitudinal modes of light that are not the gravitational wave carrying mode.
Finally, the detector strain data
is derived from the field impinging on the photo diode (PD) by means of a DC readout scheme \cite{Degallaix_2010,Hild_2009}.

The squeezed light source is
located on an external optical bench operated in-air\cite{Vahlbruch2006,Vahlbruch2010, Khalaidovski2012}.
The squeezed vacuum states of light are generated by parametric down-conversion \cite{PhysRevLett.57.2520} inside an  optical parametric amplifier (OPA) cavity, using a periodically poled potassium titanyl phosphate (PPKTP) crystal as the nonlinear medium.
The required OPA pump field at 532\,nm is generated in a second-harmonic generator (SHG).
The power of the pump field injected into the OPA can be adjusted via a Mach-Zehnder interferometer to realize different squeezing levels.  
The squeezed field is circulated into the GEO\,600 dark port via the polarizing beam splitter (PBS) of a low-loss, in-vacuum Faraday isolator. Two additional in-air Faraday isolators are used to further decouple the OPA from residual interferometer light. Two piezo-actuated steering mirrors (PZT1, PZT2 in Fig.\,\ref{fig:aufbau}) are used for automatic alignment of the squeezed field to the interferometer \cite{Schreiber2016}.

Our current work focuses on the interfacing of the squeezed light source to the detector,
in particular matching the optical mode of the squeezed light source efficiently to that of the gravitational wave signal mode of the interferometer. 
Upgrades important to improvements in the squeezing level include in-vacuum diagnostic photodiodes for in-situ monitoring of polarization matching, additional waveplates to correct for birefringence effects, and compensation of astigmatic effects in the output optics of the detector.

In the rejection ports of the in-vacuum faraday isolator (see figure~\ref{fig:aufbau}) we placed photodiodes to monitor the rejected power for understanding the polarization state of the squeezed field in the in-vacuum path \cite{SchreiberPhD}. 
We also used the signals from these PDs to optimize the PBS cube angles of incidence.

The addition of a quantum noise lock \cite{McKenzie_2005}, which actuates on the demodulation phase for the squeezer, improved the stability of the squeezed light source early on \cite{Dooley2015}.
We now use the same technique to control for drifts in the temperature of the OPA, further improving the stability of the squeezing level by maintaining the coresonance condition in the OPA.
A consistently optimized squeezed light source is important for allowing us to use the squeezing level as a metric for tuning the many degrees of freedom in the squeezing interface which impact squeezing.

To maintain consistent, very high squeezing levels, the stability of the squeezed light source in reference to the detector and it's environment as well as the interface to the detector are important.
The combination of efforts discussed in this section have improved the squeezing level from 3.4\,dB \cite{Dooley2016GEOChallenges} to the current record of 6.0\,dB which will be discussed in more detail in section~\ref{sec:results}.

\section{\label{sec:limits}Analysis}

The observed squeezing level will always be less than that of the generated squeezed vacuum due to residual optical loss, phase noise, and technical noise.
Squeezed light sources of the type employed at GEO have demonstrated very high levels of squeezing mainly limited by detection \cite{Mehmet2019}.
In GEO\,600, the squeezing level is mainly limited at high frequency by optical loss and therefore this has been our primary focus of work for achieving a high squeezing level.
In the following we discuss the loss mechanisms that affect the observable squeezing.

\paragraph*{\textbf{Optical loss -}}
Optical loss provides a path for unsqueezed vacuum to couple to the optical mode of the detector and therefore reduces the effective quantum reduction that can be achieved.
Considering a squeezed vacuum variance $V_s$ with losses, a small additional optical loss $L$ reduces the squeezing ratio by the factor,
\begin{equation}
    \sqrt{\frac{V_s}{V}} \approx 1 - \frac{V_0 - V_s}{2 V_s}L,
\end{equation}
where $V_0$ and $V$ are the variances of the unsqueezed vacuum and squeezed vacuum with additional optical loss $L$ respectively.
Thus, the stronger the squeezing, the more the loss term matters.
At the 6\,dB level of observed squeezing, a removal of an element with optical loss of 1\,\% results in an increase of detection volume of about 7\,\%.

Direct measurements of optical loss for most of the squeezed light injection path can be performed while the interferometer is not resonating and are listed in table \ref{tab:optloss}.
For this measurement, a bright field that is resonant in the OPA is injected towards the interferometer.
When the interferometer is unlocked and the end mirrors are misaligned, the reflection from the signal recycling mirror is 90\,\%.
Reflection from the locked cavity would ideally be 100\,\%, but depends on internal losses of the signal recycling cavity as well as coupling efficiencies between the relevant cavities which can be frequency dependent. Our estimate of this effective loss from the SRC is listed in table \ref{tab:optloss} as SRC loss for the measured squeezing frequency of 5.2\,kHz.

\begin{table}[b]
\caption{\label{tab:optloss}
Optical loss contributions. These are the known contributors of optical loss taking into account double passes as appropriate. Included is the equivalent loss due to dark noise from the main photodiode \cite{Grote2016} which is white at high frequency and mimicks optical loss. The total loss from this estimate is less than the total loss inferred from the observed squeezing level of about 22\,\% (see figure~\ref{fig:sqz_vs_asqz}).}
\begin{ruledtabular}
\begin{tabular}{ p{5cm} p{2cm}  }
 Source & Loss (\%) \\
 \hline
 OPA escape efficiency & 1.0 \\
 Squeezed light source after OPA & 1.3 \\
 In-air injection path & 1.8 \\
 In-vacuum optics up to OMC & 6.6 \\
 BDO1 transmission & 2.0 \\
 SRC loss & 1.4 \\
 OMC internal loss & 1.9 \\
 OMC mode matching - 2nd order & 1.3 \\
 OMC additional mismatch & 5.0 \\
 OMC alignment control dither & 0.3 \\
 PD quantum efficiency & 1.0 \\
 Dark noise equivalent & 0.1 \\
 \hline
 Total & 21.4 \\
\end{tabular}
\end{ruledtabular}
\end{table}

\begin{figure*}
\includegraphics[width=\linewidth]{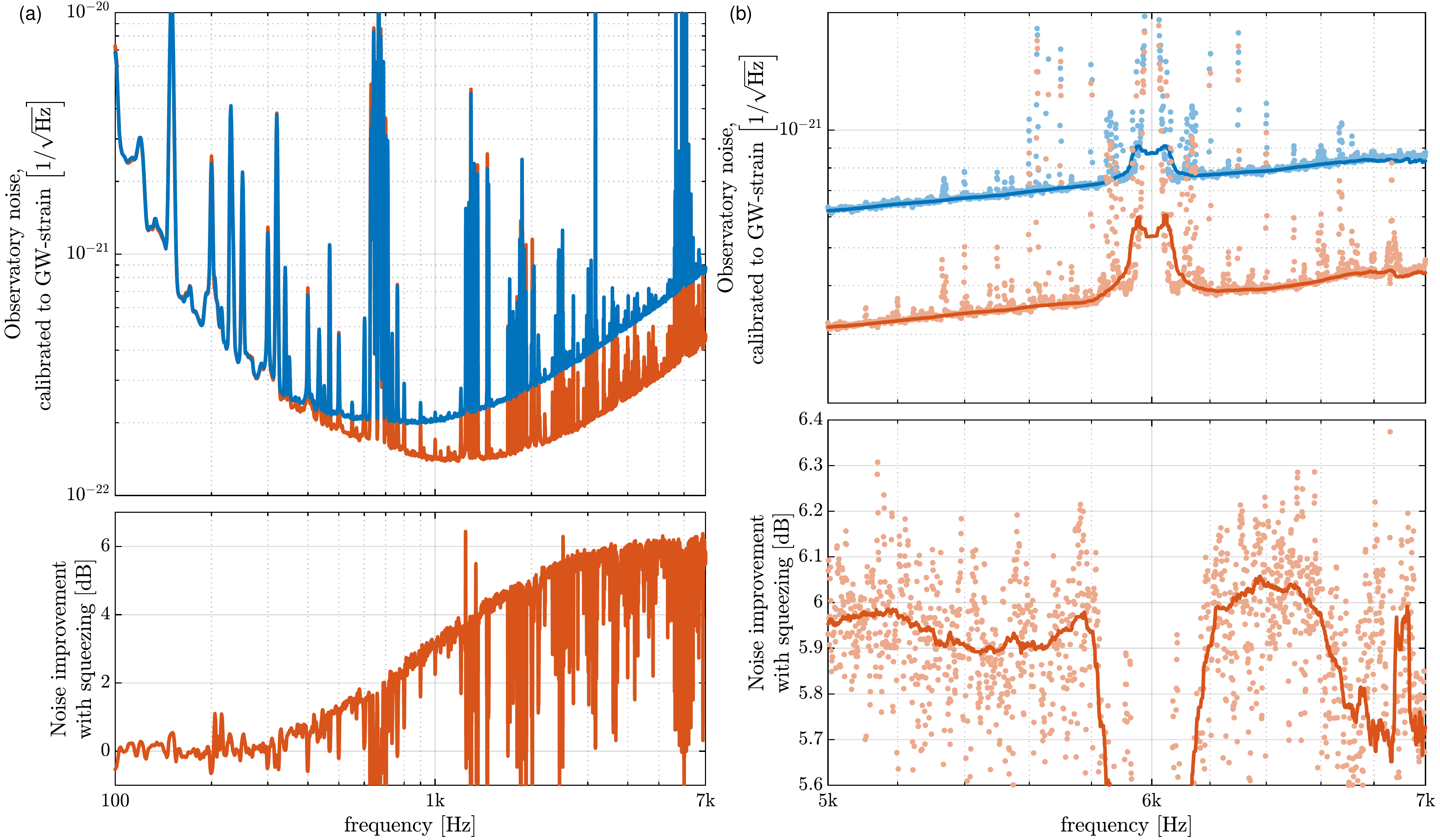}

\caption{\label{fig:sqzref} With the application of squeezing we observe a factor of two improvement in sensitivity at high frequencies.
(a) The top plot compares 15 minute power spectral density (PSD) estimates of the detector noise with (red) and without (blue) squeezing.
Towards higher frequencies, the increase in shot noise proportional to $f$ is due to calibration compensation of the filtering effect of the signal recycling cavity (SRC).
The bottom plot shows the squeezing ratio as a function of frequency.
The squeezing value of $6.03 \pm 0.02$\,dB is calculated from the ratio between median noise floor estimates with and without squeezing averaged between 6.3 and 6.5\,kHz and referenced against periods where the squeezing was not applied during valid science time over a two month duration.
Towards lower frequencies, the squeezing is increasingly limited by technical noise.
Plot (b) shows a zoomed in section of the same data (dots). Noise floor estimates in the top plot and their ratio in the bottom plot are depicted as solid lines.
}
\end{figure*}

Mode matching and alignment of the interferometer to the output mode cleaner is accomplished in reference to the so-called beacon mode of the interferometer which is generated by modulating the differential arm length at 3.17\,kHz and marks the mode which carries the gravitational wave signal.
We measure the beacon modulation at the main photo-diode and correct with the DC level to derive a new signal representing the light power of the beacon mode \cite{Smith-Lefebvre2011}.
The squeezed field is aligned to the interferometer sidebands using an autoalignment system which aligns the coherent control sidebands, which are co-propagating with the squeezed vacuum field, to the modulation sidebands from the interferometer \cite{Schreiber2016}.
Residual misalignment fluctuations will also cause optical loss. These contributions to optical loss are not expected to be very significant and are not well characterized and so are not listed in table \ref{tab:optloss}.

In order to match the mode of the squeezed vacuum field to that of the OMC, the same configuration as for optical loss measurements is used.
We achieve a matching of at least 93.8\,\%, which is determined by measuring the bright alignment light in reflection of the OMC.
Only 1.3\,\% of the mismatch is due to waist size and location, which is primarily limited by residual astigmatism.
The source of the additional 5\,\% mismatch is unknown.

\paragraph*{\textbf{Phase noise -}}
The squeezing ellipse angle is selected for maximum reduction of quantum shot noise in the readout channel. Static offset and time-dependent fluctuations - jitter - in this angle reduces the effective squeezing by coupling anti-squeezing to the readout quadrature.
The phase angle is locked using the beat of the coherent control sidebands from the OPA with the carrier field of the interferometer at the main photodiode.
Sources of phase jitter have
been published in \cite{Dooley2015}, with more recent results in \cite{SchreiberPhD} and add up to 17\,mrad.
Additionally, one can measure the impact of increased squeezing on the shot noise and use a model of the phase noise effect to estimate the phase noise.
One limitation we have found with this method is that any additional noise that is dependent on the nonlinear gain will add to the effect and give an overestimate of the phase noise.
An example of such nonlinear gain dependent noise is explained under technical noise.
The estimate of phase noise from the model fitting technique is shown in figure~\ref{fig:sqz_vs_asqz}.

\paragraph*{\textbf{Technical noise -}}
Observable squeezing is limited by technical noise which can be broadband or frequency dependant.
Technical noise in GEO generally increases towards lower frequencies and diminishes the noise improvement due to squeezing as shown in figure~\ref{fig:sqzref}.
At high frequency, the observed squeezing level is affected by two technical noises in particular that are white at the shot noise dominated frequencies in reference to the main photodiode.
These are the photodetector dark noise and a new source of technical noise we've identified as a form of backscatter noise which behaves similar to the phase noise.

The backscatter noise comes from stray interferometer light leaking along the squeezing injection path to the OPA where it is reflected with a gain factor that depends on the phase of the backscattered light. 
This reflection can cause linear and nonlinear coupling of the stray light field's random phase fluctuations $\Phi+\delta\phi$ to the detector signal, as well as a linear coupling of the residual squeezing-angle fluctuations $\delta\theta$. The backscatter noise signal is
\begin{eqnarray}
    & i_\textup{bs}(t) \approx 
      \sqrt{\frac{\eta_\textup{inj}P_\textup{stray}}{P_\textup{out}}} \nonumber \\
       \times &\left[e^{-r}\cos(\Phi) - e^{-r}\sin(\Phi)\delta\phi
      - 2\sinh(r)\sin(\Phi)\delta\theta \right]
    \,,
    \label{eq:backscatter_signal2}
\end{eqnarray}
with squeezing parameter $r$ \cite{SchreiberPhD}. The final term provides the amplification with increased squeezing. The phase noise of the squeezed light field is modulated with the absolute phase $\Phi$ of the backscattered light.
Using this effect we derive an error signal of the phase of the backscattered light by measuring the response due to a dithering of the squeezing phase.
From this signal we actuate on the path length from the squeezed light source to the interferometer with a PZT mounted mirror.

\section{\label{sec:results}Results}

\begin{figure}
    \centering
    \includegraphics[width=\linewidth]{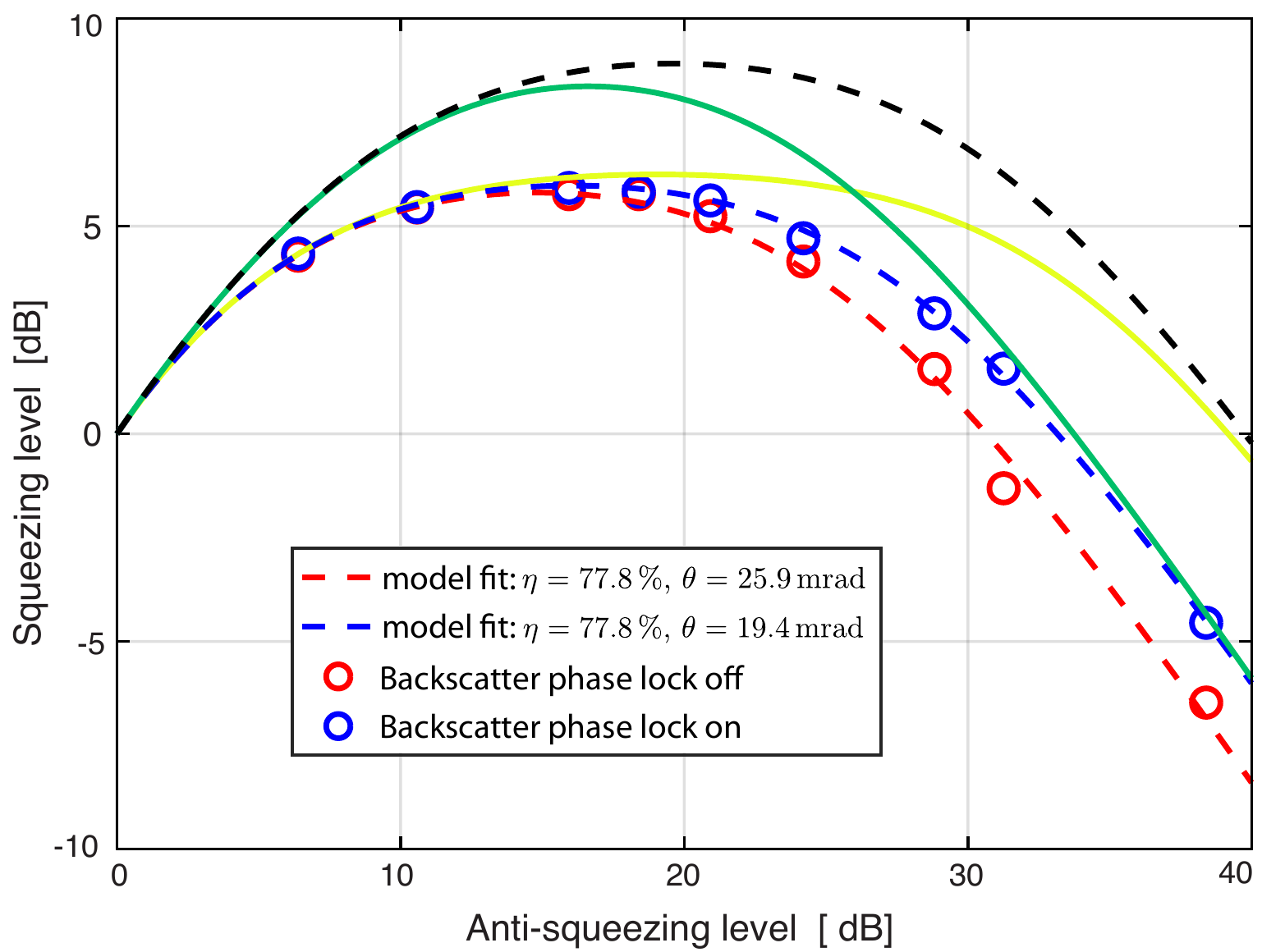}
    \caption{\label{fig:sqz_vs_asqz}Detected squeezing level as a function of measured anti-squeezing measured at 5.2\,kHz. The blue and red traces represent fits to the respective data points. The fitting procedure fits a common parameter for optical efficiency, $\eta$, 
    The two solid traces represent estimates of the detectable squeezing level if a factor of two reduction of optical loss (green) and phase noise (yellow) could be realized. 
    The dashed black trace is a factor of two improvement in both. This shows that we are still primarily limited by optical loss.}
\end{figure}

In figure~\ref{fig:sqzref}, we report up to $6.03 \pm 0.02$ dB sensitivity enhancement observed at GEO\,600 due to the application of squeezing.
We estimate the optical loss and phase noise by measuring the squeezing and anti-squeezing at different nonlinear gains in figure~\ref{fig:sqz_vs_asqz}.

Our best estimate of the sources of phase noise add up to 17\,mrad.
With the influence of backscatter noise described in seciton \ref{sec:limits}, we observe a nonlinear gain dependent noise equivalent to 25.9\,mrad of phase noise. After controlling the phase of the backscattered light field we achieve a better phase noise estimate of 19.4\,mrad. 
In figure \ref{fig:sqz_vs_asqz} we show the results of fitting our measurements to our model.
With the backscatter phase loop engaged there is a clear reduction in the estimated phase noise when fitting squeezing vs antisqueezing.

\section{Conclusions and Outlook}

The observation of 6\,dB quantum shot noise reduction from the application of squeezed states of light at the GEO\,600 detector is a first for a km-scale gravitational wave detector.
This demonstrates the ability to achieve a key goal for the next phase of upgrades to the advanced detectors known as aLIGO+ and AdV+ \cite{LIGO_IS_WP}.
The application of frequency-independent squeezing leads to an improvement in sensitivity, which can be used in addition to an increase in laser power.

This is relevant to other gravitational wave detectors given the similar complexity.
The LIGO and Virgo detectors are well on their way at 3.2\,dB~\cite{Tse2019,Acernese2019} and will face similar challenges of optical loss an mode matching in order to achieve the 6\,dB mark.
The next milestone in squeezing level to look towards is 10 dB.
This is the level of squeezing assumed for third generation detectors \cite{Punturo_2010} such as the Einstein Telescope \cite{Abernathy2011EinsteinGW} and Cosmic Explorer \cite{reitze2019cosmic}.

Optical efficiency requirements for achieving 10\,dB squeezing are better than 90\,\%, dependant on the phase noise.
For GEO\,600 we are limited primarily by optical loss, so we continue to focus in that direction towards the next goal of observing 10\,dB squeezing.
We have a few unknown losses as indicated in table~\ref{tab:optloss} as well as the discrepancy
between the measured optical loss and the observed effective optical loss that may be related to residual alignment fluctuation unaccounted for.

We are currently designing improvements to the optical system at the output of the detector, including a new OMC design.
Significant improvements in the output optics to reduce wavefront distortions and improve alignment control along with improvements to the optical efficiency of individual elements should allow for 10\,dB squeezing to be observed at the GEO detector even with the current squeezed light source.

\begin{acknowledgements}

The authors would like to thank Walter Grass for his years of expert support in the maintenance of critical infrastructure to the site to include the extensive 400\,$\mathrm{m}^3$ vacuum system.

The authors are grateful for support from the Science and Technology Facilities Council (STFC) Grant Ref: ST/L000946/1, the University of Glasgow in the UK, the Bundesministerium f\"{u}r Bildung und Forschung (BMBF), the state of Lower Saxony in Germany, the the Max Planck Society, Leibniz Universit\"at Hannover and Deutsche Forschungsgemeinschaft (DFG, German Research Foundation) under Germany's Excellence Strategy – EXC 2123 QuantumFrontiers – 390837967.
This work was partly supported by DFG grant SFB/Transregio 7 Gravitational Wave Astronomy.
This document has been assigned LIGO document number LIGO-P2000032.

\end{acknowledgements}

\bibliography{GEOsqz,Mendeley}

\end{document}